# Nonlinear photoacoustic wavefront shaping (PAWS) for single speckle-grain optical focusing in scattering media

Puxiang Lai[†], Lidai Wang[†], Jian Wei Tay[†], and Lihong V. Wang*

*Optical Imaging Laboratory, Department of Biomedical Engineering, Washington University in St. Louis, St. Louis, Missouri 63130-4899*

[†] These authors contributed equally to this work

*Corresponding author: lhwang@wustl.edu

**Abstract**

Non-invasively focusing light into strongly scattering media, such as biological tissue, is highly desirable but challenging. Recently, wavefront-shaping technologies guided by ultrasonic encoding or photoacoustic sensing have been developed to address this limitation. So far, these methods provide only acoustic diffraction-limited optical focusing. Here, we introduce nonlinear photoacoustic wavefront shaping (PAWS), which achieves optical diffraction-limited (i.e., single-speckle-grain) focusing in scattering media. We develop an efficient dual-pulse excitation approach to generate strong nonlinear photoacoustic (PA) signals based on the Grueneisen memory effect. These nonlinear PA signals are used as feedback to guide iterative wavefront optimization. By maximizing the amplitude of the nonlinear PA signal, light is effectively focused to a single optical speckle grain. Experimental results demonstrate a clear optical focus on the scale of 5–7 μm, which is ~10 times smaller than the acoustic focus in linear dimension, with an enhancement factor of ~6,000 in peak fluence. This technology has the potential to provide highly confined strong optical focus deep in tissue for microsurgery of Parkinson's disease and epilepsy or single-neuron imaging and optogenetic activation.

# Introduction

Scattering of light by wavelength-scale refractive index changes is the reason that media such as paper, frosted glass, fog, and biological tissue appear opaque[1]. The distortion of the optical wavefront propagating within such scattering media makes conventional lens focusing impossible at depths, as the optical wavelets no longer add up in phase at the targeted position. This phenomenon fundamentally limits high-resolution optical imaging techniques, such as two-photon microscopy and optical coherence tomography, to depths up to a single transport mean free path (~1 mm in soft tissue)[2]. Invasive procedures, such as embedding optical fibers, are often resorted to when concentrated light is desired beyond this depth, such as in optogenetics[3] and photothermal therapy[4]. When coherent light propagates in a scattering medium, speckles are formed. Despite appearing random, speckles are deterministic within the speckle correlation time. This property has spurred recent advances in optical time-reversal and wavefront-shaping techniques to manipulate the optical wavefront and form a focus within a scattering medium.

Optical time-reversal focusing is achieved by sensing and phase-conjugating the re-emitted wavefront from either an internal virtual guide star provided by focused ultrasound (TRUE[5-12] and TROVE[13]) or a physical guide star provided by embedded fluorescent particles[14]. In contrast, wavefront-shaping focusing is achieved by optimizing the incident wavefront to maximize the signal from a guide star. This pattern can be found using iterative algorithms[15-17], or by measuring the so-called "transmission matrix"[18]. For absorptive targets, photoacoustic (PA) sensing is preferred[19-23], as the signal comes directly from the target, as well as being non-harmful and non-invasive.

So far, focusing by PA-guided wavefront shaping has produced acoustic diffraction-limited spots. Here, we show that it is possible to beat the acoustic diffraction limit and focus light to a single optical speckle grain, reaching the optical diffraction limit. We use a novel mechanism to obtain a nonlinear PA signal based on the Grueneisen memory effect. Unlike most other nonlinear phenomena, this new mechanism produces nonlinear signals highly efficiently, enabling detection with high signal-to-noise ratio. Using this nonlinear signal as feedback, PA wavefront shaping (PAWS) achieves single speckle-grain focusing even when a large number of speckle grains are present within the acoustic focus. We demonstrate this principle and show a clear optical focus on the scale of 5–7 µm, which is ~10 times smaller than the acoustic focus, with an enhancement of peak fluence (J/m$^2$) by ~6,000 times.

## Principle

The PA effect describes the formation of acoustic waves due to absorption of light, which is usually short pulsed. The PA amplitude is proportional to the product of the absorbed optical energy density and the local Grueneisen parameter. It is well known that the Grueneisen parameters of many materials are highly temperature dependent. For example, from 25 $^{\circ}$C to 40 $^{\circ}$C, the Grueneisen parameters of water and blood can increase by 58% and 76%, respectively [2,24]. Within the thermal confinement time, the temperature rise due to the absorption of light lingers and changes the local Grueneisen parameter accordingly, which is referred to as the Grueneisen memory effect.

Here, we employ a dual-pulse excitation approach to obtain a nonlinear PA signal based on the Grueneisen memory effect. As shown in Fig. 1a, two identical laser pulses are fired sequentially to excite the same absorber. At the first laser pulse, the Grueneisen parameter is determined by the initial temperature. At the second laser pulse, the Grueneisen parameter is changed (usually increased) due to the Grueneisen memory effect. Therefore, the second PA signal has an amplitude different from the first one. If we assume that the PA amplitude is proportional to the laser energy and the Grueneisen parameter is linearly dependent on the local temperature, the amplitude difference between the two PA signals is proportional to the square of the laser energy (or fluence), yielding a nonlinear signal despite that both original PA signals are generated linearly with the current optical fluence. A detailed derivation is shown as follows.

The peak-to-peak amplitude of the first PA signal is given by the following integral:

$$V_1 = k \iint A(x,y) \Gamma_0 \mu_a F(x,y) dx dy, \tag{1}$$

where $k$ is a constant coefficient, $A(x,y)$ is the normalized acoustic detection sensitivity, $\Gamma_0$ is the Grueneisen parameter at the initial temperature $T_0$, $\mu_a$ is the material absorption coefficient, and $F(x,y)$ is the optical fluence distribution. From here on, all PA amplitudes refer to peak-to-peak values. Within the acoustic resolution voxel, both $\Gamma_0$ and $\mu_a$ are assumed to be uniform and constant, and the integration along the $z$-axis direction is taken into account in the constant coefficient $k$. $A(x,y)$ is frequently approximated using a Gaussian function, $A(x,y) = \frac{1}{2\pi w^2} \exp(-\frac{x^2+y^2}{2w^2})$, where $2\sqrt{\ln 2} w$ is the full width at half maximum (FWHM) of the one-way transducer response.

The Grueneisen parameter immediately before the second laser pulse can be approximated as

$$\Gamma = \Gamma_0 + \eta \Gamma_0' \mu_a F, \tag{2}$$

where $\eta$ is a constant coefficient that converts absorbed optical energy density into temperature rise, and $\Gamma_0^{'}$ is the first-order derivative of the Grueneisen parameter with respect to temperature at $T_0$. Therefore, the amplitude of the second PA signal is

$$V_2 = k \iint A(x,y)[\Gamma_0 + \eta \Gamma_0^{'} \mu_a F(x,y)] \mu_a F(x,y) dxdy. \quad (3)$$

The amplitude difference between the two PA signals is

$$\Delta V = V_2 - V_1 = k\eta \Gamma_0^{'} \mu_a^2 \iint A(x,y) F^2(x,y) dxdy. \quad (4)$$

This amplitude difference $\Delta V$ is determined by the square of the optical fluence, thus we term it the nonlinear PA amplitude.

When the amplitude from a single PA signal is used as feedback to iterative wavefront shaping (which we term linear PAWS), optical energy is concentrated into the acoustic focus[19-23]. To focus light to the optical diffraction limit, we use the nonlinear PA amplitude $\Delta V$ as feedback (which we term nonlinear PAWS). The reason for the narrower optical focus can be explained by rewriting equation (4) as

$$\Delta V = k\eta \Gamma_0^{'} \mu_a^2 (\overline{F}^2 + \sigma_F^2), \quad (5)$$

where $\overline{F} = \iint A(x,y) F(x,y) dxdy$ and $\sigma_F^2 = \iint A(x,y)[F(x,y) - \overline{F}]^2 dxdy$ can be treated as the mean and variance of $F(x,y)$, with a probability density function of $A(x,y)$. Since both $\overline{F}^2$ and $\sigma_F^2$ are non-negative, $\Delta V$ is maximized when both $\overline{F}^2$ and $\sigma_F^2$ are maximized. $\overline{F}^2$ is proportional to $V_1^2$ and therefore reaches its maximum when light is concentrated within the acoustic focus. Maximizing $\sigma_F^2$ is the same as maximizing the uniqueness of $F(x,y)$. Therefore, if the total optical energy is constrained, $\sigma_F^2$ is maximized when all the optical energy is focused to a single speckle grain.

Fig. 1b further explains why nonlinear PAWS can focus light to a single speckle grain using an idealized example. We simplify the ultrasonic detection sensitivity to a relatively uniform distribution within a circular focal area, and assume that the total light energy is constant and evenly distributed among the speckle grains within the acoustic focus. Let us consider two different speckle patterns $i$ and $j$: speckle pattern $i$ has multiple speckle grains within the ultrasonic focus; speckle pattern $j$ has only one speckle grain. In these two cases, the two linear PA amplitudes $V_{1i}$ and $V_{1j}$ are the same, but the two nonlinear PA amplitudes $\Delta V_i$ and $\Delta V_j$ are significantly different. Compared with speckle pattern $i$, speckle pattern $j$ concentrates light onto a smaller area and thus causes a higher temperature rise, resulting in a strong nonlinear PA signal. If all speckle grains have the same area, from equation (4), the nonlinear PA amplitude can be simply expressed as

$$\Delta V = \frac{1}{M} k\eta \Gamma'_0 \mu_a^2 A_0 \frac{E^2}{s^2}, \tag{6}$$

where $M$ is the number of speckle grains (or optical modes) within the acoustic focus, $A_0$ is the constant acoustic detection sensitivity, $E$ is the total pulse energy, and $s$ is the area of one speckle grain. Equation (6) shows that the nonlinear PA amplitude $\Delta V$ is inversely proportional to $M$, and is maximized when $M = 1$ (optical diffraction-limited focusing). The peak fluence [$\sim E/(Ms)$] is also inversely proportional to $M$. Thus the nonlinear PA amplitude is proportional to the peak fluence at constant incident laser energy. Although this conclusion is based on idealized assumptions, it is helpful for estimating the order of magnitude of the peak fluence.

**Experimental results**

Our PAWS setup is illustrated schematically in Fig. 2a. The scattering medium consists of a ground glass diffuser and a layer of optically absorbing whole blood. The incident light reflected from the SLM surface was scattered by a diffuser, generating a random speckle pattern with ~5-μm speckle grains on the blood layer. A photodiode monitored the energy of each laser pulse to compensate for the PA signals. The pulse energy on the blood layer was ~0.1 mJ, within an illuminated area of ~1 cm$^2$, which corresponded to a fluence of ~0.1 mJ·cm$^{-2}$. Initially, no nonlinear PA signals were observable even at the full energy output of the laser. In order to generate detectable nonlinear PA signals, the optical fluence needs to be sufficiently high. Therefore, to increase the optical fluence within the PA sensing region, we first conducted linear PAWS (Stage 1) before nonlinear PAWS (Stage 2), as illustrated in Fig. 2b and Supplementary Movies 1 and 2. For both stages, the SLM was divided into 192×108 independently controlled blocks, and a genetic algorithm[17,25,26] was used to optimize the phase pattern on the SLM.

In linear PAWS (Stage 1), single laser pulses were fired every 20 ms to generate the PA signals. An initial PA signal (inset of Fig. 3a), averaged over 16 traces, was recorded by displaying a random phase pattern on the SLM. As shown in Fig. 3b, the PA amplitude increased as the linear PAWS optimization proceeded, corresponding to increased optical energy within the acoustic focus[19,23]. The algorithm was terminated after 800 iterations when the improvement was less than 5% over 100 iterations; at the end, the PA amplitude increased ~60 times over the initial signal (Fig. 3a). We estimated that the fluence within the acoustic focus was increased from ~0.1 to ~6 mJ·cm$^{-2}$. The last ~250 iterations with linear PAWS showed diminishing return, as indicated by the relatively flat response toward the end.

The final phase map from Stage 1 was used as the starting point for nonlinear PAWS (Stage 2). In the nonlinear PAWS experiment, we fired a pair of pulses, separated by 40 µs (limited by the maximum laser repetition rate), well within the thermal confinement time of 189 µs (estimated based on a speckle size of ~5 µm and a thermal diffusivity of ~$1.3 \times 10^{-3}$ cm$^2 \cdot$s$^{-1}$). The initial PA signal pair, obtained by using the phase map from Stage 1, is shown in Fig. 4a. The final PA signal pair after 1600 iterations is shown in Fig. 4b, which also shows the optimized phase pattern displayed on the SLM as an inset. The enhancement of the nonlinear PA amplitude with iteration in Stage 2 is shown in Fig. 4c. The last 250 iteration improved the enhancement factor by only 5%. As seen, the final nonlinear PA amplitude was ~100 times greater than the initial value, indicating a ~100-time improvement of the peak fluence. To avoid overheating the blood during the optimization, the laser energy was attenuated by 10% every 300 iterations. At the beginning of each adjustment, $\Delta V$ was re-measured. All other parameters were kept constant. The change in energy was compensated for in the results shown in Figs. 4b and 4c. The nonlinear signal plateaued toward the end of the optimization, indicating that the focal spot had approached its smallest size.

We imaged the optical field at the ultrasonic focal plane using a CCD camera. When a random phase pattern was displayed on the SLM, a speckle pattern (Fig. 5a) was captured with randomly distributed speckle grains. The FWHM of the acoustic focus is shown by the dashed circle. Note that there are many speckle grains within the acoustic focus. When the optimized phase pattern from nonlinear PAWS was displayed, a focal spot with the size of a single speckle grain was formed (Fig. 5b). The size of the focal spot was measured to be 5.1 μm × 7.1 μm (FWHM), which is ~10 times smaller than that of the acoustic focus.

## Discussion

So far, optical focusing using PA-guided wavefront shaping has been limited by acoustic diffraction when extended optical absorbers are targeted. To break through the acoustic resolution limit, we have proposed and demonstrated nonlinear PAWS. Using dual-pulse excitation, nonlinear PA signals were generated based on the Grueneisen memory effect. By maximizing the nonlinear PA amplitude, we were able to focus diffuse light into a single optical speckle grain. The focus was measured to be 5.1 μm × 7.1 μm, about an order of magnitude smaller than the acoustic focal size in linear dimension. Note that, about 169 speckle grains existed within the acoustic focal region (estimated by taking the ratio between the area of the acoustic focus and the area of a single speckle grain), but after nonlinear PAWS, only one became dominant. While the experiments here were conducted with an absorber positioned behind a ground glass diffuser, optical focusing inside a scattering medium would be similar, as long as the medium is nearly acoustically transparent as in soft biological tissue.

The peak fluence enhancement was estimated to be ~6000 times, ~60 times from the linear PAWS stage and ~100 times from the nonlinear PAWS stage. The peak fluence enhancement can also be estimated from the temperature rise. At the end the nonlinear PAWS, the second PA amplitude $V_2$ was ~168% greater than the first PA amplitude $V_1$, which was measured at a room temperature of 25°C (Fig. 4b). Assuming that the Grueneisen parameter of blood is proportional to the temperature rise[24], we estimate the corresponding temperature rise to be ~33°C. Note that instantaneous (submilliseconds) temperature rises of this magnitude does not cause biological damages[27]. From here, we predict the final fluence $F$ as

$$F = \frac{\Delta T \rho C_V}{\mu_a} = \frac{33\,\text{K} \times 1\,\text{g}\cdot\text{cm}^{-3} \times 3600\,\text{mJ}\cdot\text{g}^{-1}\cdot\text{K}^{-1}}{240\,\text{cm}^{-1}} = 495\,\text{mJ}\cdot\text{cm}^{-2}, \qquad (7)$$

where $\rho$ is the mass density of blood, $C_V$ is the heat capacitance of blood, and $\mu_a$ is the absorption coefficient of blood. Compared to the initial fluence of ~0.1 mJ·cm$^{-2}$, the final peak fluence is increased by ~4950 times, which agrees with the aforementioned estimation of ~6000 times.

While most nonlinear phenomena are weak, photoacoustic nonlinearity based on the Grueneisen memory effect is exceptionally strong, primarily due to the dependence of the thermal expansion coefficient with temperature[28]. As shown in Fig. 4b, the nonlinear signal $\Delta V$ was even stronger than the first linear signal $V_1$. It is worth noting that both of the original PA signals are produced linearly with the current incident laser fluence. This strong PA phenomenon as observed using dual-pulse excitation based on the Grueneisen memory effect will likely find broad applications in biomedical optics.

To date, there has been only one other demonstration of noninvasive speckle-scale optical focusing inside scattering media, by using time reversal of variance-encoded light (TROVE)[13]. In TROVE, the scattered light is recorded with multiple illumination speckle realizations while a focused ultrasound beam is used to define the target region. Speckle-scale focusing is then obtained by computing the appropriate phase map from the measured speckle fields. Despite achieving similar goals, TROVE and nonlinear PAWS are complementary: TROVE time-reverses ultrasonically encoded light, and is therefore more applicable for non- or low-absorption targets. In comparison, nonlinear PAWS is preferred in applications with optically absorptive targets, such as blood vessels or melanomas in biological tissue. Furthermore, the peak enhancement reported in TROVE is ~110 with digital background subtraction, whereas we have demonstrated an unprecedented peak enhancement of ~6,000 without background subtraction.

Such an orders-of-magnitude peak enhancement with a well-defined virtual guide star can potentially advance many laser applications in deep tissues, such as laser microsurgery and single-neuron optogenetic activation, that benefit from intense and highly confined focusing. Examples of microsurgery include photocoagulation of small blood vessels and photoablation of tissue[29]. Without invasive probes, laser microsurgery is limited to depths of several hundred micrometers[30]. The peak enhancement by nonlinear PAWS can be used to extend the operating depths while single speckle grain focusing is maintained. This type of microsurgery could potentially lead to new noninvasive treatment of Parkinson's disease or epilepsy. Similarly, nonlinear PAWS can potentially enable deep imaging of the brain at single-neuron resolution.

Our current setup can be improved to increase the optimization speed. Linear and nonlinear PAWS currently take several hours in total. To maintain the deterministic property of the scattering medium, the PAWS focusing procedure must be completed within the speckle correlation time, which is on the order of one millisecond for *in vivo* tissue. We are currently limited by the slow response of the SLM used. Although the SLM can operate at 60 Hz, we have found that it takes about 1.2 s for the SLM display to fully stabilize[26]. Due to this long optimization time, we demonstrated the principle using a stable diffuser. In the future, faster devices can be used to accelerate the optimization. For example, digital mirror devices with switching times of 22 μs have been used in wavefront shaping[31], and could shorten the optimization. The speed also affects our choice in the number of controlled blocks used on the SLM. On one hand, the optimization time with the genetic algorithm scales linearly with the number of blocks[17,26]. On the other hand, the potential peak enhancement also increases linearly. We chose to use $192 \times 108$ as a practical compromise.

The generation of nonlinear PA signal requires only a moderate instantaneous (rather than continuous) temperature rise. We used an initial fluence of 6 mJ·cm$^{-2}$ for nonlinear PAWS, which is well below the ANSI safety limit of 20 mJ·cm$^{-2}$.[32] To avoid potential thermal damage, the laser energy was attenuated during the nonlinear optimization. On one hand, since nonlinear PAWS successfully proceeded with fluence as low as 6 mJ·cm$^{-2}$, the laser energy could be further reduced. On the other hand, the high optical fluence after nonlinear PAWS could be leveraged for laser microsurgery at optical resolution deep in tissue.

In this work, we assume that the nonlinear PA signal is quadratic with the laser pulse energy, based on the linear temperature dependence of the Grueneisen parameter. However, even in the presence of higher-order effects, nonlinear PAWS can still lead to optical diffraction-limited focusing. It should also be noted that the optical focal spot produced using nonlinear PAWS is near the center of the acoustic focus. However, the precision is limited by the signal-to-noise ratios of the final PA signals and the exact acoustic focal profile.

The expected peak improvement factor for phase-only (i.e., no amplitude optimization) wavefront-shaping is given by[11,33]

$$\text{Factor} = \frac{\pi}{4}\frac{N-1}{M}+1, \tag{8}$$

where $N$ is the number of independently controlled SLM blocks, which was 192×108 in our study, and $M$ is the number of optical speckle grains (i.e., optical modes) within the acoustic focus, which was ~169 in the linear PAWS stage. Thus, the theoretical enhancement ratio from the linear PAWS was 97. Experimentally, we measured an enhancement of ~60 (Fig. 2b). The difference could be due to the laser-mode fluctuation, non-uniformity of optical illumination on the SLM, stray light, mechanical instability of the system, and measurement errors. Nonetheless, after linear PAWS, the optical fluence within the acoustic focus was sufficient to generate detectable nonlinear PA signals. After nonlinear PAWS, the number of bright speckle grains should ideally reduce from ~169 to 1. Hence, we expected an improvement factor of ~169 after nonlinear PAWS. In the experiment, the improvement was ~100 (Fig. 4c). The less than expected performance was probably due to the same factors affecting linear PAWS. The peak fluence enhancement of ~6,000 is also approximately consistent with the expected improvement factor from equation (8) when $M$ after nonlinear PAWS was reduced to ~2-3, counting the "residual" darker speckle grains in Fig. 5b.

In closing, we have demonstrated a nonlinear PAWS approach to break the acoustic resolution limit and achieve both optical resolution focusing and a high peak-enhancement factor in scattering media. While the present study was performed using whole blood as the absorbing target, the Grueneisen memory effect exists broadly in many materials[34]. Therefore, similar performance can be anticipated with other types of absorbers. Furthermore, the system can conceivably be engineered to respond much faster. Nonlinear PAWS opens an avenue for many micrometer-scale optical applications, including imaging, sensing, therapy, and manipulation, deep inside highly scattering biological tissue.

## Methods

**Experimental setup**. The experimental setup is schematically illustrated in Fig. 2a, with more details shown in Supplementary Fig. S1. We used a 532 nm pulsed laser (INNOSLAB BX2II-E, EdgeWave GmbH, Germany), which produced 10 ns pulses (pulse energy ≤ 0.2 mJ) at an adjustable pulse repetition rate of 0–30 kHz. The laser beam was directed through a half-wave plate and a polarizing beam splitter to sample a small fraction of the beam. Light reflected by the beam splitter was attenuated and measured using a photodiode (PDA36A, Thorlabs, USA), and was used to compensate for energy fluctuations of the laser output. Light transmitted by the beam splitter was expanded, and then reflected off a liquid-crystal-on-silicon (LCoS) based phase-only SLM (PLUTO, Holoeye Photonics, Germany). The SLM had an aperture of 16 mm by 9 mm, with a resolution of 1920×1080 pixels. In the experiment, the SLM was evenly divided into 192 × 108 blocks, each independently controlled, with a linearized[35] phase shift between 0 and $2\pi$. The reflected beam was condensed using a set of lenses, and focused by a microscopic objective (10X, NA=0.25) onto a ground glass diffuser (DG10-120, Thorlabs, USA; the turbidity of the diffuser is illustrated in Supplementary Figure S2). A neutral density filter wheel between the SLM and the objective lens reduced the laser fluence in nonlinear PAWS experiments to avoid thermal saturation. A circular container (15 mm diameter, 4 mm height) of bovine blood was placed 10 mm away from the diffuser to serve as the absorptive target for PA sensing. A focused ultrasonic transducer (homemade based on a non-focusing transducer; more details below) was positioned on the other side of the blood layer to detect the PA signal. Both the blood layer and ultrasonic transducer were immersed in water for acoustic coupling. The water was maintained at room temperature by circulation.

**Detection of PA signals and control of optimization**. The PA signals generated were amplified by 50 dB (ZFL-500LN+ and ZX60-43-S+, Mini-Circuits, USA), digitized by an oscilloscope (TDS5034, Tektronix, USA) at a bandwidth larger than 500 MHz, and sent to a computer. The linear and nonlinear PA amplitudes were quantified in MATLAB (R2012b, MathWorks, USA), and a genetic algorithm[17,25,26] controlled the optimization. The phase map was displayed on the SLM using a graphics card (GeForce GT520, NVidia, USA). A digital delay generator (DG645, Stanford Research Systems, USA) controlled the synchronization between the laser and the oscilloscope. For linear PAWS, one pulse was fired every 20 ms. For nonlinear PAWS, two pulses were fired with a delay of 40 μs, but the burst period remained at 20 ms. After the optimization, the blood layer was moved off the optical path, and a CCD camera attached to a microscope—with a resolution of 1 μm/pixel—was used to image the optical field at the ultrasound focal plane (Fig. 5), when the initial and final phase patterns were displayed on the SLM, respectively. By calculating the autocorrelation[36] of the initial speckle pattern, we measured the speckle grain size at the ultrasonic focal plane to be ~5 μm, which was consistent with the final experimental optical focus size.

**Transducer field calibration**. A 50-MHz focused ultrasonic transducer was used in the experiment. The transducer was modified in-house from a non-focusing transducer (V358, Panametrics NDT, USA) by adding an acoustic focusing lens. Due to the high center frequency, the typical method of characterizing the transducer using a hydrophone or a pulser-receiver cannot be used. Instead, we used acoustic phase conjugation from a metal ball (8 mm diameter)[37,38] to measure the acoustic focal zone. The transducer axial focus was measured to be 11.425 mm from the transducer, and the lateral FWHM of the focal region was 65 μm. See Supplementary Fig. S2 for more details.

## Acknowledgements

The authors thank Konstantin Maslov for manufacturing the acoustic lens, Cheng Ma for assistance on the acoustic focus calibration, Tsz-Wai Wong for help on preparing the supplementary cartoons, and James Ballard for editing the manuscript. This work was sponsored in part by National Institute of Health grants DP1 EB016986 (NIH Director's Pioneer Award) and R01 CA186567 (NIH Director's Transformative Research Award) as well as National Academies Keck Futures Initiative grant IS 13.

## Author Contributions

P.L., J.W.T., and L.V.W. initiated the project. P.L. implemented the photoacoustic wavefront shaping system. L.W. initiated the principle of dual-pulse PA nonlinearity based on the Grueneisen memory effect. J.W.T. wrote code for the experiment and simulations. P.L., J.W.T., and L.W. designed and ran the experiment, and prepared the manuscript. L.V.W. provided overall supervision. All authors were involved in the analysis of the results and manuscript revision.

## Competing financial statement

P.L., J.W.T., and L.W. declare no competing financial interests. L. V. W. has financial interests in Microphotoacoustics, Inc. and Endra, Inc., which, however, did not support this work.

## Additional information

Supplementary information is available in the online version of the paper. Correspondence and requests for material should be addressed to L.V.W.

**Figure captions**

**Figure 1. Principles. a**, Illustration of dual-pulse excitation producing a nonlinear photoacoustic signal based on the Grueneisen memory effect. Two laser pulses with equal energy E are incident on an optical absorber. The first pulse causes a lingering change in the Grueneisen parameter—referred to as the Grueneisen memory effect—due to an increase in temperature. Within the thermal confinement time, the change in the Grueneisen parameter $\Delta\Gamma$ causes the amplitude from the second PA signal ($V_2$) to be stronger than that from the first ($V_1$). The difference between the peak-to-peak amplitudes $\Delta V$ is nonlinear—proportional to the square of the laser pulse energy (or fluence). **b**, Illustration of nonlinear PAWS principle. When the same optical energy is concentrated to fewer speckle grains within an acoustic focus, the linear PA amplitude does not increase significantly, but the nonlinear PA amplitude approximately increases inversely proportionally with the number of bright speckle grains. Therefore, by maximizing $\Delta V$, light can be focused as tightly as the optical diffraction limit (i.e., one speckle grain). The blue dashed circles represent the ultrasonic focal region.

**Figure 2. Experimental setup and dual-stage optimization. a**, Schematic of the photoacoustic wavefront shaping (PAWS) experimental setup. PBS, polarized beam splitter; SLM, spatial light modulator; $\lambda/2$, half-wave plate. **b**, Illustration of the two-stage optimization procedure (see Supplementary Movies 1 and 2 for more information). Stage 1, linear PAWS focuses light into the acoustic focal region. Stage 2, nonlinear PAWS focuses light onto a single-speckle grain. The blue dashed circles represent the acoustic focal region. A typical intensity distribution (green solid line) is shown above the speckle illustrations. The blue dashed envelopes represent the acoustic sensitivity.

**Figure 3. Experimental results of Stage 1—using linear PA signal as feedback for wavefront shaping (linear PAWS)**. **a**, PA signals before (blue dashed curve) and after (red solid curve) the linear PAWS (Stage 1) optimization. Note that all PA signals in this study were compensated for laser energy fluctuations, and normalized to the initial PA peak-to-peak amplitude shown here. **b**, Linear improvement factor (defined as the ratio of the PA amplitudes to the initial PA amplitude) versus iteration index. Linear PA amplitude improved ~60 times in Stage 1, indicating a peak enhancement factor of ~60 for optical fluence within the acoustic focus.

**Figure 4. Experimental results of Stage 2—using nonlinear PA signal as feedback for wavefront shaping (nonlinear PAWS)**. **a**, The initial PA signal pair (blue dashed curve for the first, and red solid curve for the second) from the paired laser pulses (separated by 40 µs) in Stage 2, when the phase pattern obtained from the linear PAWS was displayed on the SLM. The difference between the two PA signal amplitudes ΔV was used as feedback in nonlinear PAWS. **b**, The final PA signal pair (blue dashed curve for the first, and red solid curve for the second) after Stage 2 optimization. Note that the shown signals have been compensated for the laser pulse energy adjustments shown in **c**. The first PA signal remained relatively constant before and after nonlinear PAWS, but the second PA signal was significantly enhanced because of the Grueneisen memory effect. The inset shows the final optimized phase pattern displayed on the SLM. **c**,

Nonlinear improvement factor (defined by the ratio of the compensated nonlinear PA amplitudes to the initial value) versus iteration index. Nonlinear PA amplitude improved ~100 times during Stage 2. 1600 iterations were used, with the incident laser energy ($E$) attenuated by 10% every 300 iterations to avoid overheating the sample. The normalized laser energy $R = E/E_{max}$ is also shown, where $E_{max}$ was the initial laser energy used before adjustment. The compensated nonlinear PA amplitudes are given by $\Delta V / R^2$, and the nonlinear improvement factor is therefore given by $\frac{\Delta V / R^2}{\Delta V_{initial}}$, where $\Delta V_{initial}$ denotes the initial $\Delta V$.

**Figure 5. Visualization of single speckle grain focusing using nonlinear PAWS. a**, Speckle pattern observed behind the diffuser when a randomized phase pattern was displayed on the SLM. **b**, Optical focus down to a single speckle grain observed behind the diffuser when the optimized phase pattern from Stage 2 (the inset of Figure 4b) was displayed on the SLM. The 1D profiles across the focus (green solid curves) measure 5.1 and 7.1 μm along *x* and *y*, respectively. The blue dashed circles show the measured acoustic focal region (50 MHz, −6 dB). Its lateral profiles (blue dashed curves) measure a FWHM of 65 μm. The intensity values in **a** and **b** are normalized to their own peak values.

# Figure 1

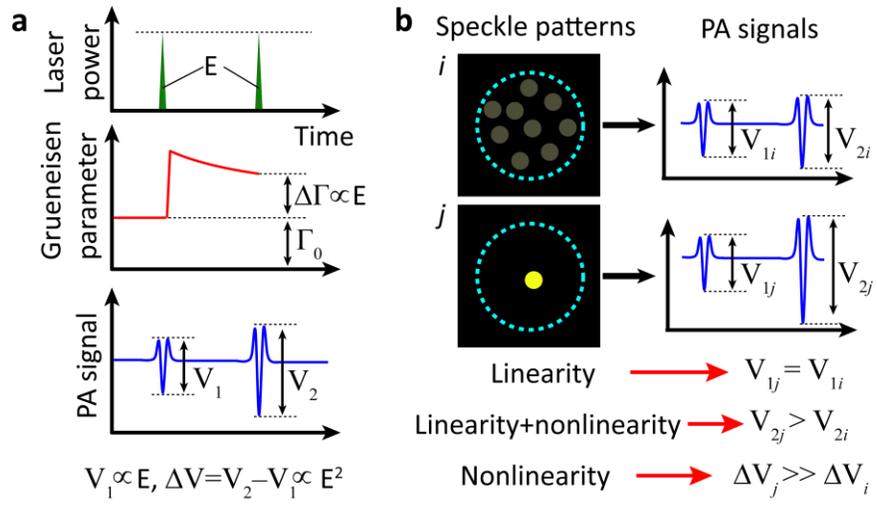

**Figure 2**

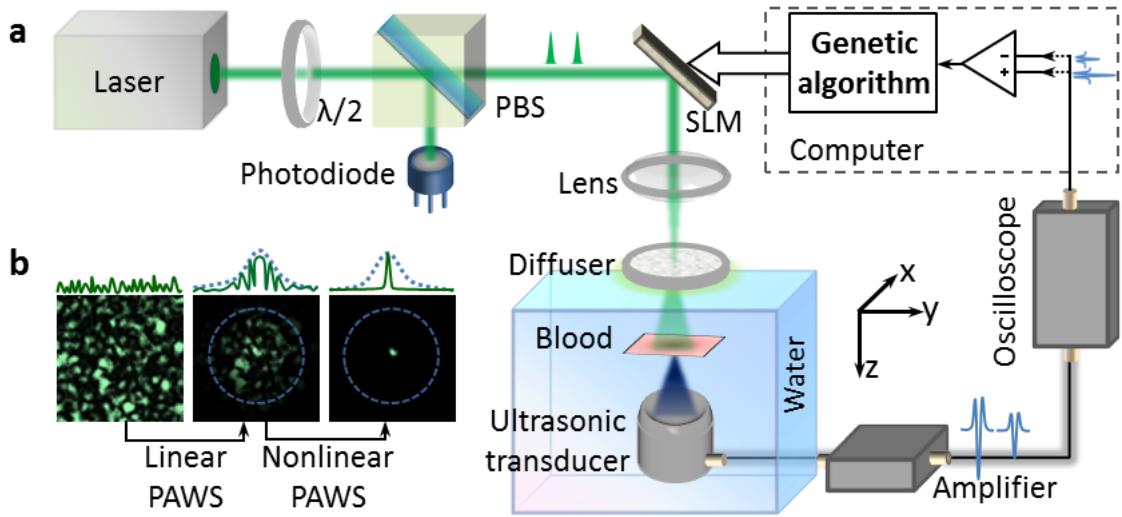

**Figure 3**

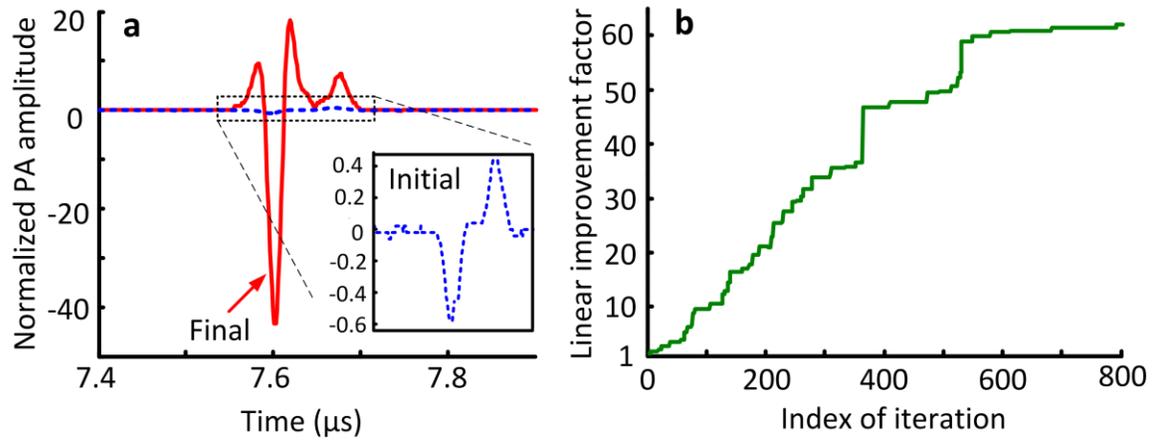

**Figure 4**

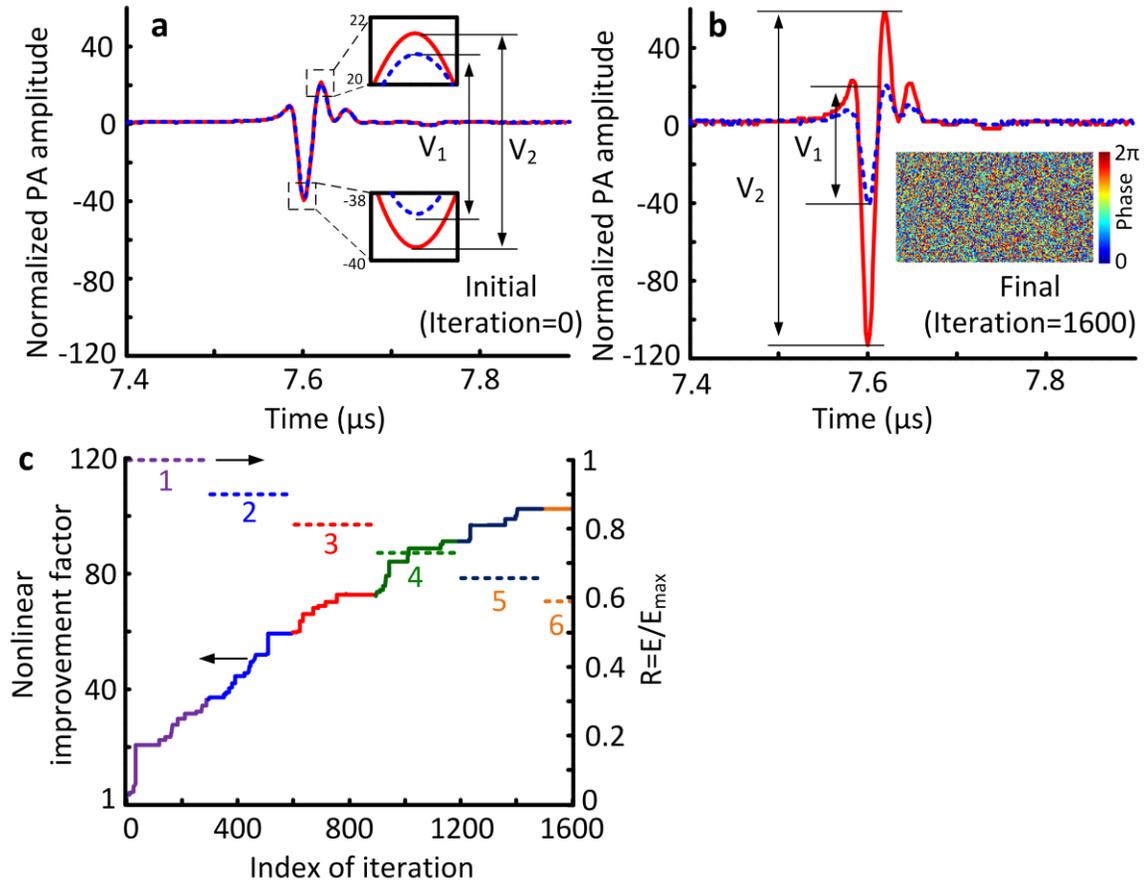

**Figure 5**

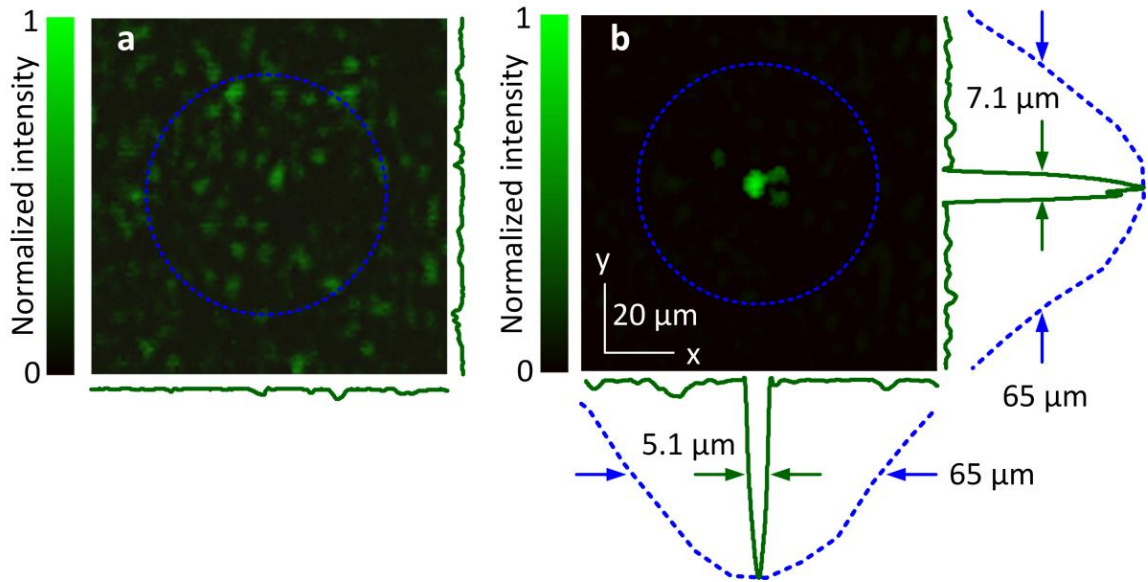